# Microstructured Optical Waveguide-Based Endoscopic Probe Coated with Silica Submicron Particles

**Timur Ermatov** [1], **Yury V. Petrov** [2], **Sergei V. German** [1], **Anastasia A. Zanishevskaya** [3], **Andrey A. Shuvalov** [3], **Vsevolod Atkin** [2], **Andrey Zakharevich** [2], **Boris N. Khlebtsov** [2,4], **Julia S. Skibina** [3], **Pavel Ginzburg** [5], **Roman E. Noskov** [5], **Valery V. Tuchin** [2,6,7] and **Dmitry A. Gorin** [1,*]

1. Skolkovo Institute of Science and Technology, 3 Nobelya str., 121205 Moscow, Russia; timur.ermatov@skolkovotech.ru (T.E.); s.german@skoltech.ru (S.V.G.)
2. Saratov State University, 83 Astrakhanskaya str., 410012 Saratov, Russia; welcometospain@mail.ru (Y.V.P.); ceba91@list.ru (V.A.); lab-15@mail.ru (A.Z.); khlebtsov_b@ibppm.ru (B.N.K.); tuchinvv@mail.ru (V.V.T.)
3. SPE LLC Nanostructured Glass Technology,101 50 Let Oktjabrja, 410033, Saratov, Russia; zan-anastasiya@yandex.ru (A.A.Z.); shuvalovaa@nano-galss.ru (A.A.S.); skibinajs@yandex.ru (J.S.S.)
4. Institute of Biochemistry and Physiology of Plants and Microorganisms, 13 Prospekt Entuziastov, 410049 Saratov, Russia
5. Department of Electrical Engineering, Tel Aviv University, Ramat Aviv, Tel Aviv 69978, Israel; pginzburg@post.tau.ac.il (P.G.); nanometa@gmail.com (R.E.N.)
6. Tomsk State University, 36 Lenin's av., Tomsk 634050, Russia
7. Institute of Precision Mechanics and Control of the Russian Academy of Sciences, 24 Rabochaya str., 410028 Saratov, Russia
\* Correspondence: d.gorin@skoltech.ru; Tel.: +79172077630

**Abstract:** Microstructured optical waveguides (MOW) are of great interest for chemical and biological sensing. Due to the high overlap between a guiding light mode and an analyte filling of one or several fiber capillaries, such systems are able to provide strong sensitivity with respect to variations in the refractive index and the thickness of filling materials. Here, we introduce a novel type of functionalized MOWs whose capillaries are coated by a layer-by-layer (LBL) approach, enabling the alternate deposition of silica particles ($SiO_2$) at different diameters—300 nm, 420 nm, and 900 nm—and layers of poly(diallyldimethylammonium chloride) (PDDA). We demonstrate up to three covering bilayers consisting of 300-nm silica particles. Modifications in the MOW transmission spectrum induced by coating are measured and analyzed. The proposed technique of MOW functionalization allows one to reach novel sensing capabilities, including an increase in the effective sensing area and the provision of a convenient scaffold for the attachment of long molecules such as proteins.

## 1. Introduction

Microstructured optical waveguides (MOWs) have matured into a major area of cutting-edge science since its inception in 1996 [1]. They have found numerous applications in optics and related fields [2], including the production and modification of solar cells [3], biosensors [4–6], biomedical investigation [7–9], endoscopy [10,11], and clinical imaging [12,13]. In combination with graded-index (GRIN) lenses [14,15], MOWs have been employed in optical microscopy [10,16,17] and microendoscopy [18]. Hollow capillaries of MOWs (HC-MOWs) filled with fluids offer a powerful playground for in-fiber microfluidic optical sensing, enabling measurements of fluid's refractive index, temperature, fluorescence signals, or biochemical agent concentrations [19]. They are also very prospective for neurophotonics studies and applications, specifically, in the field of optogenetics and monitoring/controlling of the blood-brain barrier [20–26].

Approaches for liquid sensing within MOWs can be subdivided into two categories: the first relies on detecting optical properties of molecules via light absorption [27], fluorescence [28], or Raman scattering [29]. In the solid-core MOWs, light propagates in the core, and the mode evanescent tail probes the optical properties of an analyte in nearby holes.

The second category of MOW-based sensors exploits fiber resonant features that can be very sensitive to variations in the refractive index within the holes. Such resonant features can be caused by either Bragg or long-period gratings [30,31], intermodal interferences, e.g., in tapers [32], surface plasmon resonances [33,34], or the photonic bandgap properties of MOWs themselves [35,36]. HC-MOWs belong to the latter class. Here, the analyte fills one or several fiber capillaries, serving as a part of the probed resonator, and the optical resonances of these fluid-filled channels are probed. The major benefit of using MOWs rather than equivalent techniques based on cuvettes and bulk optics lies in combining the long interaction lengths with strong overlapping between light and the analyte within small sample volumes, giving rise to high sensitivity [37–40].

Importantly, all the previous works on this topic have demonstrated sensing functionalities of MOWs for capillaries with smooth walls [41,42]. The potential of MOWs with porous walls remained obscure. In this work, we develop the technique for HC-MOWs coating with chemically and mechanically stable [43–48] and monodisperse submicron silica particles by the layer-by-layer (LBL) deposition approach [49–54]. We demonstrate up to three covering bilayers consisting of 300-nm, 420-nm, and 900-nm silica particles [55] and poly(diallyldimethylammonium chloride) (PDDA) between them. Submicron silica particles were chosen based on their optical properties and easy, highly monodisperse synthesis. Due to the refractive indexes of silica and the fiber glass being very close, they may provide the porosity of the capillary walls without too substantial rise in the transmission losses, and the monolayer of silica particles can be considered as an effective increase in the capillary wall thickness [56–59]. Modifications in the MOW transmission spectrum induced by coating are measured and analyzed. The proposed technique of MOW functionalization allows one to reach novel sensing capabilities, including an increase in the effective sensing area, providing a convenient scaffold for the attachment of long molecules such as proteins [60], and combining in-fiber liquid sensing [61–63] and high-performance liquid chromatography [64]. The surfaces, which are covered with closely packed spherical particles, can also serve as promising sensitive elements of gas sensors [65] due to capillary condensation in the gap formed with particles in contact [66–71].

## 2. Materials and Methods

### 2.1. MOW Samples and SiO$_2$ Particles

We use the MOW containing three concentric capillary layers surrounding the central hollow core (Figure 1a) [72]. The wall thickness for the capillaries of the first layer is 1.76 μm. Supermonodisperse silica nanoparticles were prepared by using the multistep method described in Ref [55] (Figures 1b, 1c, 1d). Briefly, for the synthesis of initial 24-nm silica seeds, 9.1 mg of L-arginine was added to 6.9 mL of water under magnetic stirring. Further, 0.55 mL of tetraethyl orthosilicate (TEOS) was slowly added, and the mixture was allowed to react for 20 h at 60 °C. For the synthesis of 45-nm and 68-nm silica seeds, we used a similar regrowth process, which was repeated twice. Final silica nanoparticles were regrown to a desirable size according to the modified Stöber process by varying the volume of TEOS and keeping the constant ratio between the reagent volumes: ethanol/water/ammonia/68 nm seeds = 18:2:1:1. For example, for the preparation of 300-nm silica nanoparticles, 7.6 mL of aqua ammonia was mixed with 136.8 mL of absolute ethanol and 15.2 mL of water under magnetic stirring at room temperature; this was followed by the addition of 7.6 mL of 68-nm silica seeds. At the end, 15 mL of TEOS was added over 5 h by using a syringe pump. Finally, nanoparticles were centrifuged twice at 3000 g during 30 min and resuspended in water. The final concentration for all samples of the silica particles was equal to 12.8 mg/mL.

Poly(diallyldimethylammonium chloride) (PDDA, MW = 400,000–500,000), poly(styrenesulfonate) (PSS, MW = 70,000) and polyethylenimine (PEI, MW = 2,000,000) were purchased from Sigma Aldrich (St. Louis, MO, USA). Deionized water was produced by Millipore Mili-Q Plus 185 (manufacturer, city, country).

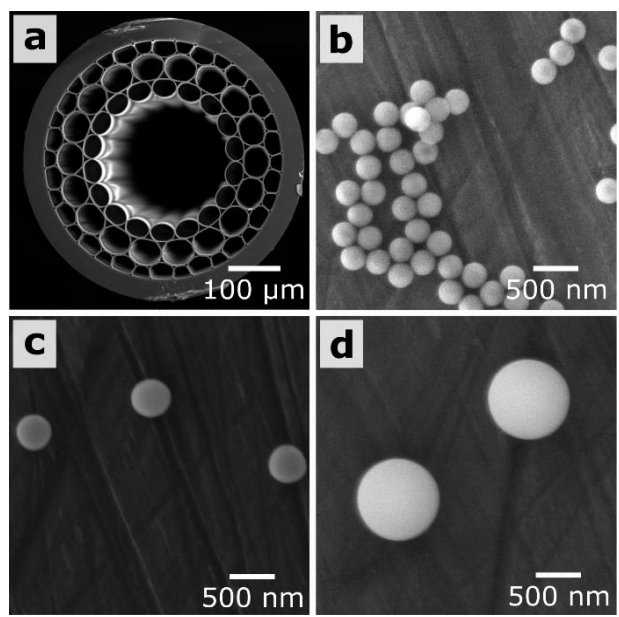

**Figure 1.** (**a**) Micrographs of scanning electronic microscopy (SEM) (manufacturer, city, country) for the unfilled microstructured optical waveguides (MOW) end face with a hollow core region of 240 μm and **(b–d)** $SiO_2$ spherical particles with diameters of 300 nm, 420 nm and 900 nm, respectively.

*2.2. Deposition Process*

We use 6 cm-long MOW samples connected to the Shenchen peristaltic pump by a flexible silicon tube with an inner diameter of 1 mm. To fix the samples inside the tubes, we produced special three-dimensional (3D) printed clamps, which ensured the efficient flow of solution through the samples. Our system supplies a highly controllable and persistent flow rate for any given solution capacity, allowing the uniform deposition of polyelectrolyte layers inside the capillaries (Figure 2).

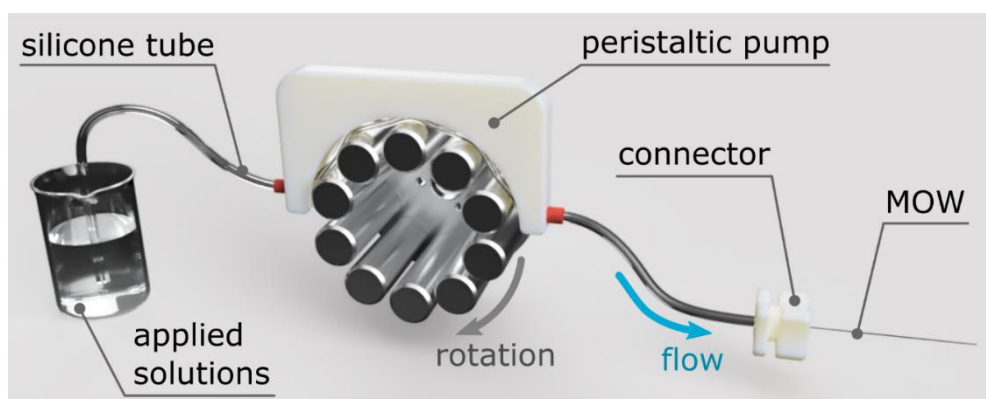

**Figure 2.** Schematic of the experimental setup for fiber coating.

In the beginning, MOW samples were washed with deionized water for 5 min with a speed of 500 μL/min to clean them from small dust particles; then, they were immersed with the LBL technique by a combination of oppositely charged polyelectrolytes (PEI/PSS/PDDA) with a concentration of 2 mg/mL in 0.15 M of water solution of NaCl during the 7 min for each layer; the pump flow rate was reduced to 100 μL/min. Deionized water was applied after each polyelectrolyte layer to wash the samples and remove unabsorbed particles. At the end, we deposited the combination of five polyelectrolyte bilayers of PSS/PDDA on the inner walls of HC-MOW (the very first layer consisted of three polyelectrolytes: PEI/PSS/PDDA), which served as a substrate (buffer layer) for the stronger adhesion of the silica particles. The more polyelectrolyte layer thickness, the greater the number of $SiO_2$ particles that adsorbed onto it through the dipper penetration into the formed polyelectrolyte substrate layer that prevents the removal of silica particles while washing MOW samples.

The possibility of the deposition of multiple silica layers was showed for 300-nm $SiO_2$ particles. One, two, and three layers of silica particles were applied on the polyelectrolyte substrate. The effect of the polyelectrolyte intermediate coating on 300-nm silica particles deposition was investigated on the example of comparison of a single PDDA layer and PDDA/PSS/PDDA triple-layer applied between silica layers.

*2.3. Optical Characterization*

Optical measurements of MOW samples were performed by the transmission setup based on a Thorlabs CCS200 spectrometer (Thorlabs, Newton, NJ, USA) and a broadband light source: a Thorlabs SLS201L halogen lamp. The incident light was launched into a MOW sample by the 4× Olympus objective (Olympus, Tokyo, Japan), and the output signal was collected by the 10× Olympus objective, and then sent to a spectrometer for spectral analysis (Figure 3). Smooth function was applied to the measured transmission spectra. The spectra before and after the application of the smooth function were presented in Figure A1 in the Appendix.

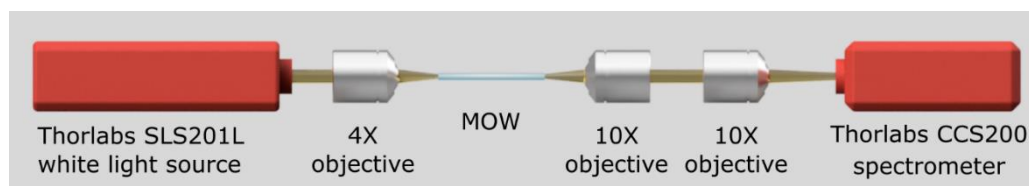

**Figure 3.** Schematics of the transmission measurement setup.

## 3. Results and Discussion

*3.1. Polyelectrolyte Buffer Layer Deposition*

LBL deposition is based on the alternative adsorption of oppositely charged polyelectrolytes. The very first layer includes three polyelectrolytes: PEI/PSS/PDDA. Then, four PSS/PDDA bilayers were applied to finalize the formation of the buffer layer, which is aimed at strong adhesion for $SiO_2$ particles. Figure 1 shows the resulted $SiO_2$ particles with diameters of 300 nm, 420 nm, and 900 nm, and the MOW end face. Following References [73–80], we can assume that the total thickness for this buffer layer consisting of five PSS/PDDA bilayers was 30 nm (3 nm per a single polyelectrolyte layer). The deposited silica particles are strongly attached to capillary walls covered by such a thick substrate layer. They are not removed by the fast water flow and stayed inside even after MOWs washing with a flow rate of 7 mL/min (the maximal value available for our microfluidic pump shown in Figure 2).

The transmission spectra of the resulted MOW sample are shown in Figure 4. The light guiding mechanism in this type of MOW can be described via Fabry–Perot resonances [38,80], and the condition corresponding to the maximal decoupling of the core and cladding modes can be written as follows:

$$\lambda_j = \frac{4 n_1 d}{2j+1}\left(\frac{n_2^2}{n_1^2}-1\right)^{1/2}, \qquad (1)$$

where *j* is an integer $(j = 0,1,2…)$, describing the mode order, $n_1$ is the refractive index of a medium filling the capillaries (which is air in our case), $n_2$ is the refractive index of the fiber glass, and *d* denotes the wall thickness for the first capillary layer. However, we notice that this model has restrictions. Specifically, it cannot provide the correct quantitative description for fiber losses and modal content away from Fabry-Perot resonances. Therefore, in the cases when these properties are relevant, the inhibited coupling model is more preferable [81].

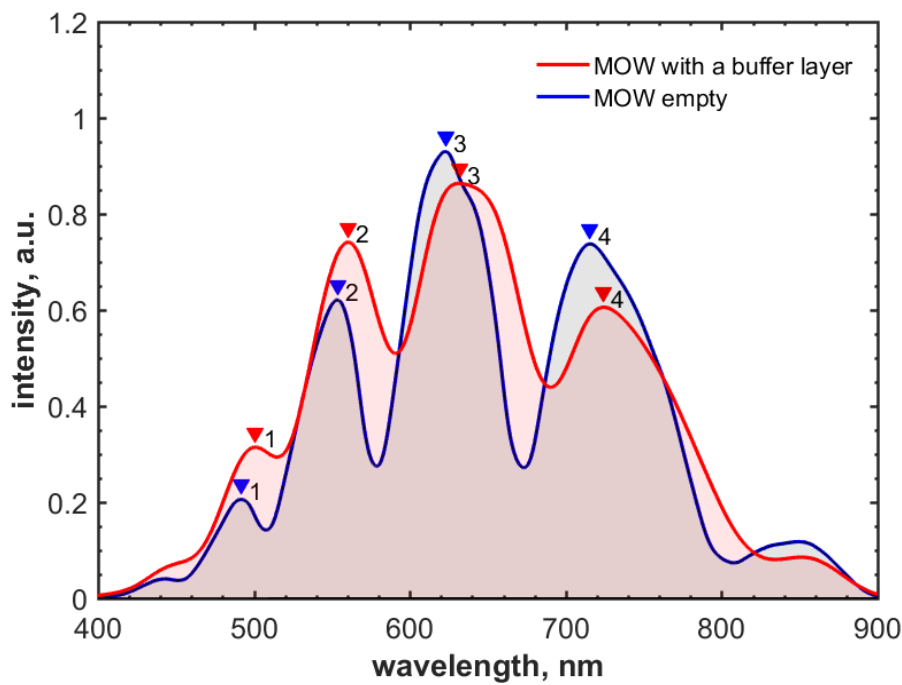

**Figure 4.** Spectral shifts of local transmission bands induced by poly(styrenesulfonate)/poly(diallyldimethylammonium chloride) (PSS/PDDA) polyelectrolyte layers. Transmission spectra for the empty (uncoated) MOW (blue) and the MOWs coated with the buffer layer (red).

Equation (1) gives the local bands of the transmission shown in Figure 4 for $j$ = 5, 6, 7, 8. The red-shift of the local transmission bands is caused by the effective increase in $d$ owing to the buffer layer (the difference in the refractive indexes of the fiber glass and polymers is negligible [58,59,82,83]). Table 1 shows excellent agreement between local band spectral shifts that are measured and predicted by Equation (1), for a total polyelectrolyte layer thickness of 30 nm.

**Table 1.** Spectral shift of local transmission bands for MOW samples modified by PSS/PDDA layers.

| Band Number | Spectral Shift, nm Measured | Spectral Shift, nm Calculated |
|---|---|---|
| Peak1 | 9 ± 2 | 8 |
| Peak2 | 10 ± 2 | 9 |
| Peak3 | 11 ± 2 | 11 |
| Peak4 | 11 ± 2 | 13 |

*3.2. Comparison of 300-nm and 420-nm Silica Particles Coating*

Next, we consider MOW samples with a deposited layer of 300-nm and 420-nm silica particles, as presented in Figure 5. Importantly, in both cases, the nanoparticles homogeneously cover almost the full surface of the core capillary, which proves our concept of the polyelectrolyte buffer layer formation for the better adsorption of $SiO_2$ nanoparticles.

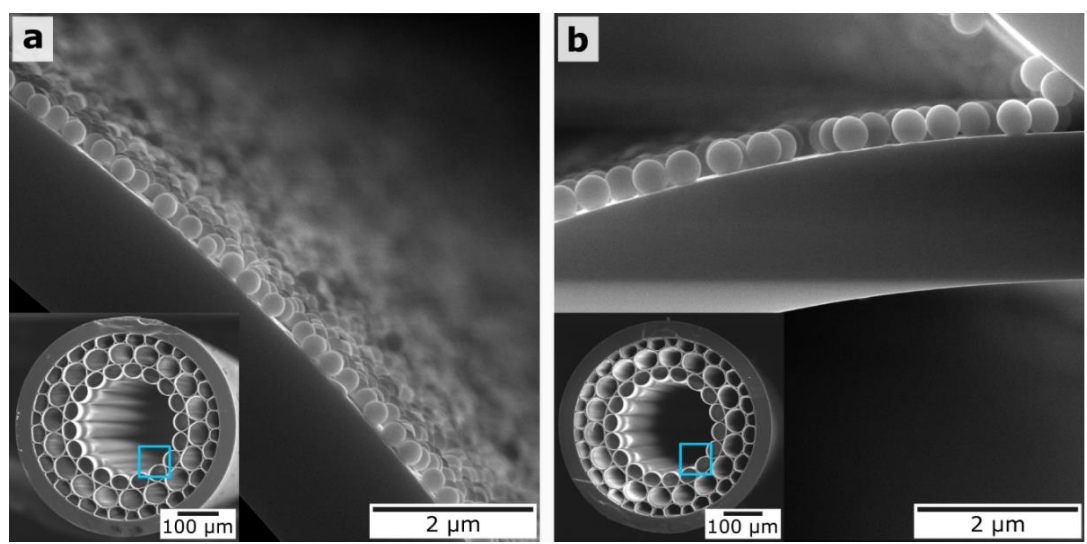

**Figure 5.** SEM images of MOW end faces and magnified capillary surfaces. (**a**) MOW sample after deposition of one layer of 300-nm silica particles. (**b**) MOW sample after deposition of one layer of 420-nm silica particles.

Figure 6 exhibits the transmission spectra of MOWs coated with the buffer layer only and 300-nm and 420-nm silica nanoparticles. The deposition time and flow rate were equal for $SiO_2$ particles of both sizes. Experimentally measured spectral shifts are in good agreement with the ones calculated from Equation (1) for the deposition of 300-nm and 420-nm silica particles, and the effective thickness of applied layers of 106 nm and 125 nm, respectively (Table 2).

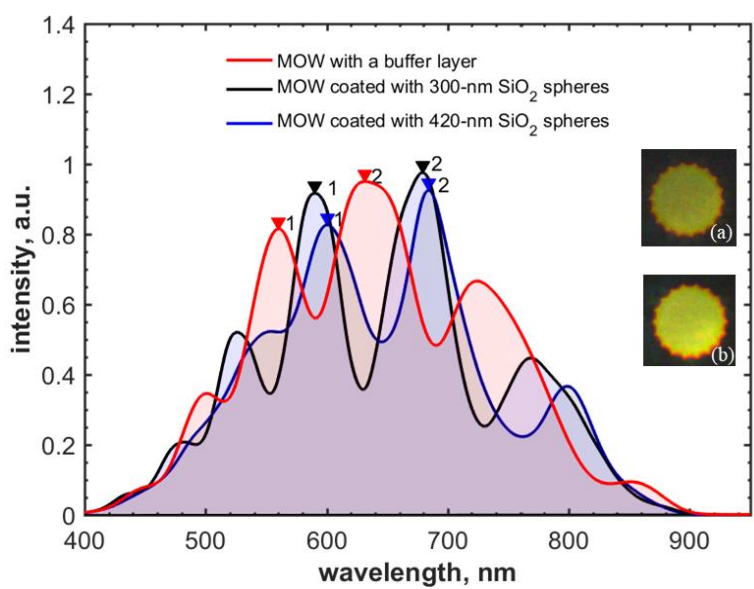

**Figure 6.** Spectral shift of local transmission bands for MOW samples with one deposited layer of 300-nm and 420-nm $SiO_2$ particles into MOWs with an internal buffer layer. Inserts are the mode profiles of MOW samples modified by (**a**) 300-nm and (**b**) 420-nm silica particles.

**Table 2.** Transmission peaks shift of MOWs with a single layer of 300-nm and 420-nm $SiO_2$ particles.

| | Spectral Shift Measured, nm | | Spectral Shift Calculated, nm | |
|---|---|---|---|---|
| Peak number | 300-nm $SiO_2$ | 420-nm $SiO_2$ | 300-nm $SiO_2$ | 420-nm $SiO_2$ |
| peak1 | 31 ± 2 | 38 ± 2 | 33 | 39 |
| peak2 | 39 ± 2 | 45 ± 2 | 38 | 45 |

From SEM images of MOWs covered by 300-nm and 420-nm silica particles (Figure 5), one can see that the capillaries became almost fully coated by $SiO_2$ particles, regardless of the total amount of 300-nm silica particles being greater than the 420-nm ones at the same volume of applied solution that came from different mass of 300-nm and 420-nm silica particles. Based on this, it can be concluded that the use of $SiO_2$ particles with a concentration of 12.8 mg/mL corresponds to a case of excess concentration for their deposition on 6 cm-long MOW samples, and the reason of the larger spectral shift induced by the coating of 420-nm $SiO_2$ particles is the larger effective thickness of the formed silica layer inside the MOW HC region than the layer thickness built from 300-nm silica particles.

*3.3. 900-nm Silica Particles Deposition*

It is instructive to note that the further growth of silica nanoparticle size leads to the deterioration of the shape of the MOW transmission curve as a result of strong light scattering. For example, we covered MOW by a suspension of 900-nm silica nanoparticles (Figure 7), which led to the significant deformation of the spectral curve (Figure 8) [84].

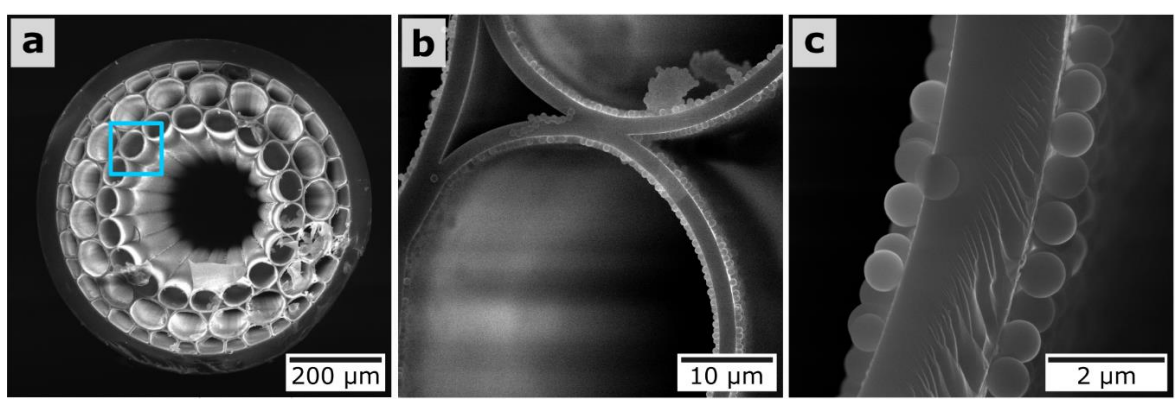

**Figure 7.** SEM images of (**a**) the MOW end face and (**b**,**c**) capillaries with a layer of 900-nm silica particles.

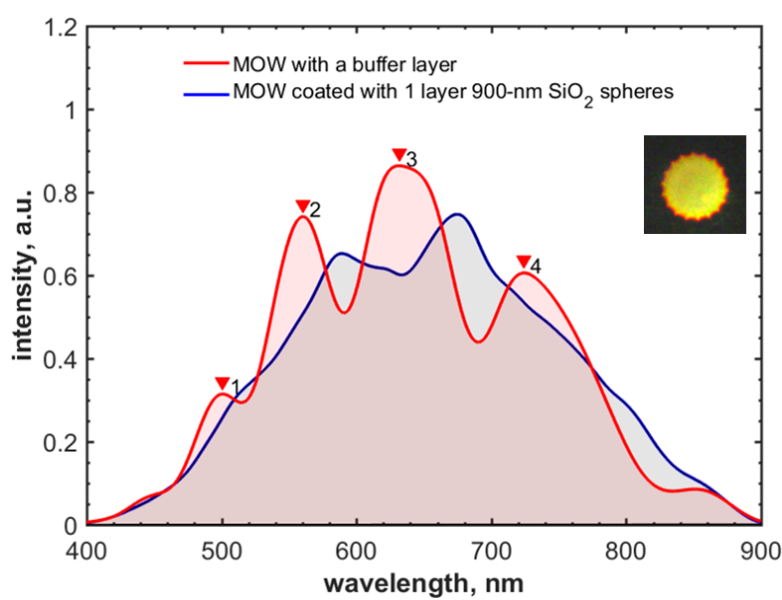

**Figure 8.** Transmission spectrum of MOW sample with one deposited layer of 900-nm silica particles (blue). MOW spectrum with a buffer layer (red). Insert shows the mode profile of the MOW coated with 900-nm silica particles.

To show the effect of the particle size on the transmission of MOWs, we measured an output power of MOW samples coated by one layer of silica particles with diameters of 300 nm, 420 nm, and 900 nm. The coupling conditions were the same (input power 19 µW). As expected, the transmitted power was decreasing as the particle size growths: 3.4 µW for 300-nm particles, 1.9 µW for 420-nm particles, and 1.2 µW for 900-nm silica particles.

*3.4. Fluid Dynamics Inside the Capillaries*

The particle concentration in the coating layer is similar to previous cases. However, in contrast to smaller particles of 300 nm and 420 nm, the bigger silica particles of 900 nm form layers on both sides of the capillary. This can be explained by Stoke's law and Bernoulli's principle. The viscosity force **F**, which describes the flow of particles inside the capillaries, is proportional to their size and velocity:

$$\mathbf{F} = 6\pi\mu R \upsilon, \quad (2)$$

where $\mu$ is the dynamic viscosity, R is particle radius, $\upsilon$ is the flow velocity.

Bernoulli's principle, which describes the volumetric flow rate, says that the fluid (particle suspension) inside the smaller capillaries will travel faster than the ones inside the bigger capillaries (Figure 9):

$$Q = A_1 \upsilon_1 = A_2 \upsilon_2, \quad (3)$$

where Q is the volumetric flow rate, $A_1$ and $A_2$ are the cross-sections of bigger and smaller capillaries, and $\upsilon_1$ and $\upsilon_2$ are the flow velocities inside the bigger and smaller capillaries, respectively.

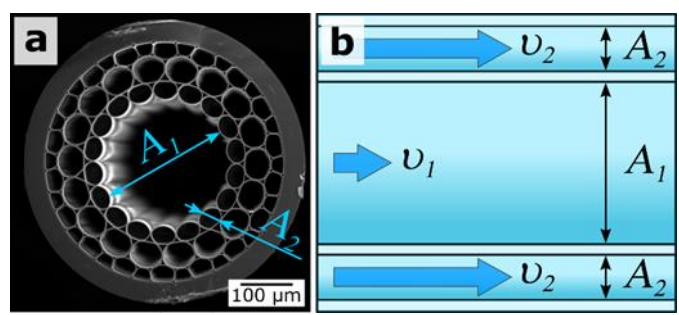

**Figure 9.** (**a**) SEM image of the MOW end face and (**b**) a schematic view of hollow core region surrounded by the smaller capillaries in the course of particle suspension deposition, $A_1$ and $A_2$ are the cross-sections of bigger and smaller capillaries, and $\upsilon_1$ and $\upsilon_2$ are the flow velocities inside the bigger and smaller capillaries, respectively.

The bigger silica particles are affected by the stronger viscosity force (Equation (2)), leading to slowing them down and, as a result, there being a higher chance of their adsorption onto capillary walls. Different local flow velocities inside large and small capillaries have the same effect on the deposition of particles (Equations (2) and (3)). The probability of particle adsorption onto the inner central hollow core surface is higher in comparison with the surrounded capillaries where the particles move faster (Equation (3)). This is well-illustrated by the strong difference between the deposition of small (300-nm and 420-nm) and large (900-nm) silica particles. In the first case, particles adhere onto the internal surface of the central capillary only; in the second, particles additionally adsorb to the first layer of capillaries (Figure 7). Such double coating results in the disappearance of the Fabry–Perot anti-resonances, which are associated with local bands in the transmission spectra (Figure 8).

*3.5. 300-nm Silica Particles Multiple Layers Coating*

Next, we analyze MOW coating with three layers of 300-nm silica nanoparticles (Figure 10). It should be noted that silica particle adsorption to a buffer polyelectrolyte layer is more efficient than the deposition of the second and the third SiO$_2$ particle layers. An intermediate PDDA layer applied between silica particles serves as a recharger and aims at their better adsorption. However, due to its small thickness (in the range of 3 nm) [73–80], a lower number of silica particles are adsorbed. This effect is illustrated in Figure 11, where the highest spectral shift was measured for the first layer of deposited silica particles (Table 3).

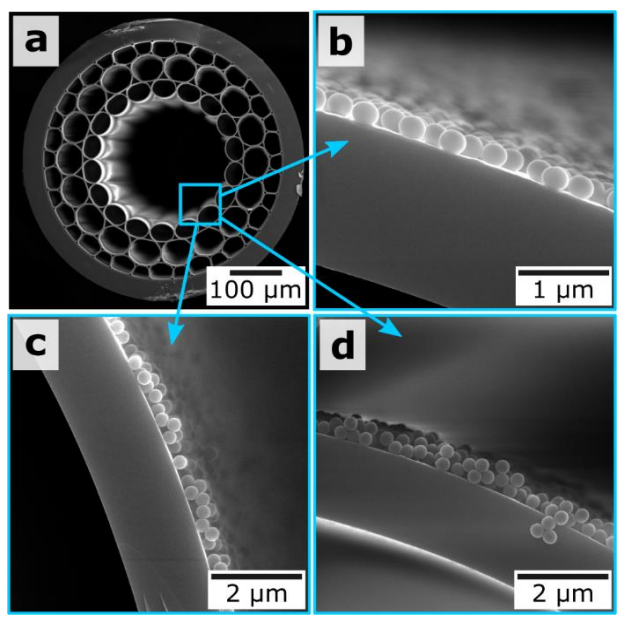

**Figure 10.** (**a**) SEM image of the MOW end face; (**b**–**d**) magnified SEM images of the capillaries with one, two, and three deposited layers of 300-nm silica particles, respectively.

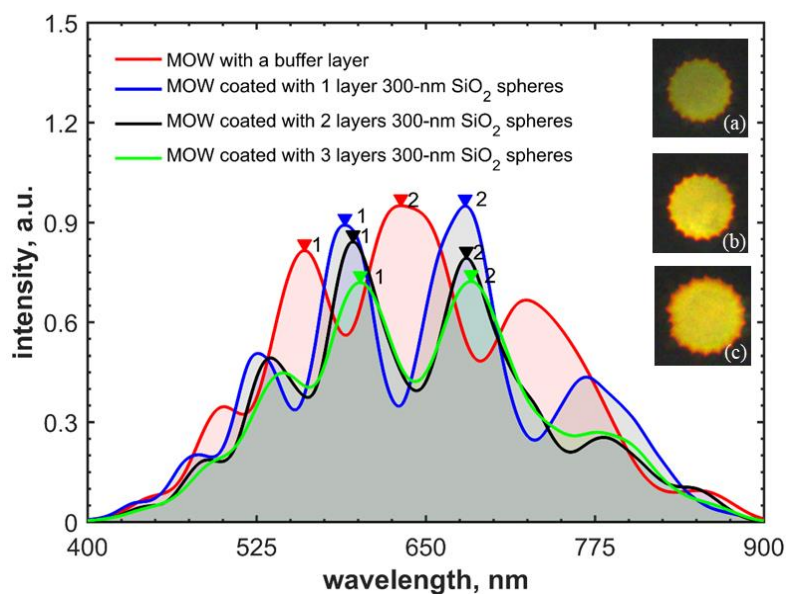

**Figure 11.** Transmission spectra for MOWs coated with the buffer layer only (red) and with one (blue), two (black), and three (green) layers of 300-nm SiO$_2$ particles. Inserts are the mode profiles of MOWs with one (**a**), two (**b**), and three (**c**) layers of 300-nm silica particles.

**Table 3.** Shifts of the transmission peaks for MOWs driven by layers of 300-nm silica particles.

| | Spectral Shift, nm | | |
|---|---|---|---|
| Band number | 1 layer | 2 layers | 3 layers |
| peak1 | 31 ± 2 | 37 ± 2 | 42 ± 2 |
| peak2 | 39 ± 2 | 42 ± 2 | 46 ± 2 |

It was established that the local transmission bands of the MOW sample with one deposited layer of 420-nm silica particles and the sample with two deposited layers of 300-nm particles were equally shifted because the product of the effective thicknesses and the refractive index of such formed silica layers were the same.

*3.6. Intermediate Polyelectrolyte Layer Effect on Silica Particles Adsorption*

To increase the concentration of adsorbed silica particles for multilayer deposition, we studied the effect of intermediate polyelectrolyte layers and compared the two cases: a single PDDA layer and a combination of PDDA/PSS/PDDA applied between $SiO_2$ particles. The triple polyelectrolyte layer was chosen to form the necessary thickness for better silica particle adsorption. The higher concentration of $SiO_2$ particles deposited on the core surface is shown in Figure 12, and the higher spectral shifts are illustrated in Figure 13 and Table 4 for two deposited layers of 300-nm silica particles.

SEM images of MOW hollow core regions proved our concept of the formation of silica layers and showed the higher particle concentration for two deposited layers with a triple layer of PDDA/PSS/PDDA than the silica concentration with a single PDDA layer deposited (Figure 12).

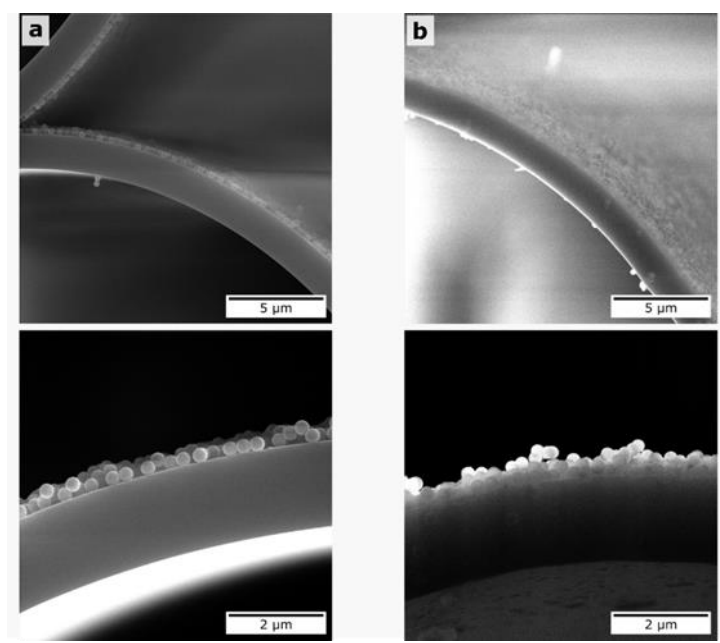

**Figure 12.** SEM images of MOW capillaries with two layers of 300-nm $SiO_2$ nanoparticles for the cases of (**a**) PDDA/PSS/PDDA and (**b**) PDDA intermediate layers, respectively.

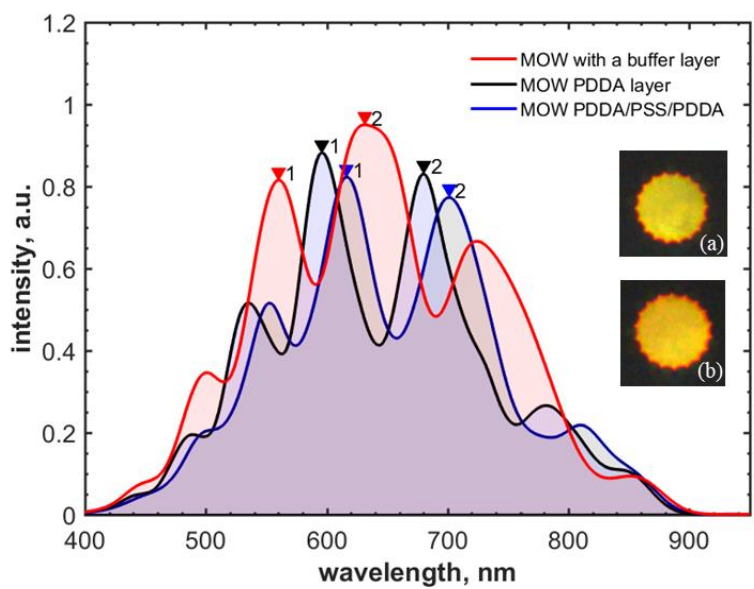

**Figure 13.** Transmission spectra for MOWs coated with the buffer layer only (red) and with two layers of 300-nm silica particles over PDDA (black) and PDDA/PSS/PDDA (blue) intermediate layers. The inserts are the mode profiles of MOW samples modified by two layers of 300-nm silica particles over PDDA (**a**) and PDDA/PSS/PDDA (**b**) intermediate layers.

**Table 4.** Spectral shift of local transmission bands for two layers of 300-nm silica particles over PDDA or PDDA/PSS/PDDA intermediate layers.

| Spectral Shift, nm | | |
|---|---|---|
| Band number | PDDA | PDDA/PSS/PDDA |
| peak1 | 37 ± 2 | 57 ± 2 |
| peak2 | 42 ± 2 | 66 ± 2 |

## 4. Conclusions

For the first time to our knowledge, the LBL technique for MOW modification through the deposition of silica particles with different sizes (300, 420, and 900 nm) onto capillary surfaces has been presented. We have analyzed the effect of such functionalization on the optical transmission of MOWs and demonstrated a clear red-shift of the transmission windows, which is increased with the rise of the particles' size and the number of deposited particles layers. However, comparatively large particles of 900 nm deteriorate the fiber transmission windows due to the strong roughness of the coating surface and the undesirable light scattering which results.

We have also shown the fiber coating with one, two, and three layers of 300-nm silica particles. The largest spectral shift in the transmission is induced by the first deposited silica layer (31 nm and 39 nm, for the first and the second transmission bands, respectively), while additional silica layers are worse-adsorbed on the first layer of particles, resulting in the additional transmission spectral shifts (11 nm and 7 nm).

The impact of the polyelectrolyte intermediate coating between the silica particle layers on their adsorption efficiency has been studied. A single PDDA layer was compared with a triplet of PDDA/PSS/PDDA. The SEM images of MOW samples coated by two silica particle layers show the higher concentration of adsorbed particles for PDDA/PSS/PDDA layers than for the case of a single PDDA layer deposition. Thus, silica particles deposition onto the formed triple-polyelectrolyte layers is more efficient.

To conclude, MOWs coated with layers of silica particles provide a convenient scaffold for the attachment of long molecules such as proteins and gas condensation. This in combination with the strong sensitivity of MOWs' transmission windows to the presence of analytes allows one to offer this type of functionalized MOW as a promising chemical and biological sensor. Specifically, this type of MOWs provides an opportunity for the static and dynamic detection of biomolecules applying the same technique as in Ref. [60], showing a selective cancer protein detection, but with an advantage of a larger effective sensing area. Beyond that, the controllable tuning of local transmittance bands and the enhancement of their contrast can be useful for laser light delivery and detection in neurophotonics for optogenetic studies and for monitoring/controlling of the blood–brain barrier by the endoscopic probe approach.

**Author Contributions:** T.E. performed the experiments; J.S.S., A.A.Z. and A.A.S. fabricated the MOW samples; B.N.K. synthesized the silica particles; T.E., D.A.G., R.E.N., S.V.G., J.S.S., B.N.K., P.G. and Y.V.P. worked on the analysis of the data and wrote the paper; D.A.G., J.S.S, and V.V.T. engaged in project administration and critical revising of the manuscript for important intellectual content; S.V.G. engaged in visualization and images preparation; V.A. and A.Z. acquired the SEM data; R.E.N. and D.A.G. supervised the projects.

**Funding:** This work was supported by the Russian Foundation for Basic Research (RFBR grant 18-29-08046). Valery V. Tuchin was supported by the Ministry of Education and Science of the Russian Federation (grant 17.1223.2017/AP).

**Conflicts of Interest:** The authors declare no conflict of interest.

**Appendix A**

The procedure of graphs preparation is shown in Figure A1 on the example of the transmission spectra of a MOW sample coated with a polyelectrolyte buffer layer and the sample with one deposited layer of 300-nm silica particles. The effect of smooth function can be seen. It does not change the general trend of the curves but makes them easier to analyze and work with.

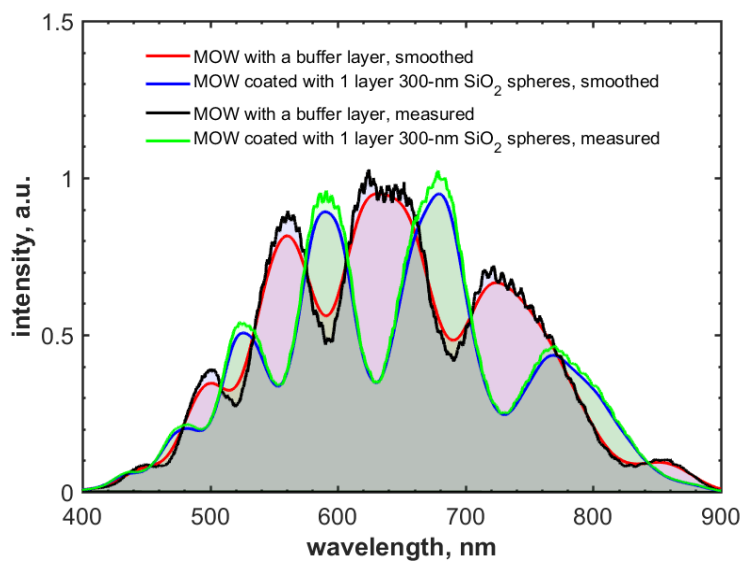

**Figure A1.** Transmission spectra for MOWs coated with the buffer layer only: measured (black) and smoothed (red). Transmission spectra for MOWs coated with one layer of 300-nm silica particles: measured (green) and smoothed (blue).